\documentclass{elsarticle}
\usepackage[justification=centering]{caption}
\usepackage{graphicx}
\graphicspath{{images/}}
\usepackage{bm}
\usepackage{tabularx}
\usepackage{amsmath}
\usepackage{amssymb}
\DeclareMathOperator*{\E}{\mathbb{E}}
\usepackage{cleveref}
\usepackage{algorithm}
\usepackage{algorithmic}

\usepackage{multirow}
\usepackage{amsmath,amsfonts,amssymb,bm}   %% AMS mathematics macros

\makeatletter
\def\hlinewd#1{
\noalign{\ifnum0=`}\fi\hrule \@height #1 \futurelet
\reserved@a\@xhline}
\makeatother
\newcolumntype{?}{!{\vrule width 0.9pt}}
\newcolumntype{~}{!{\vrule width 0.7pt}}

\title{mREAL-GAN: Generating Multiple Residential Electrical Appliance Load Profiles with Inter-Dependencies using a Generative Adversarial Network}
\author{Edward Sanderson}\corref{cor1}
\ead{esanderson4@uclan.ac.uk}
\author{Aikaterini Fragaki}
\ead{afragaki@uclan.ac.uk}
\author{Jules Simo}
\ead{jsimo@uclan.ac.uk}
\author{Bogdan J. Matuszewski}
\ead{bmatuszewski1@uclan.ac.uk}
\date{October 2021}

\address{School of Engineering, University of Central Lancashire, Preston, Lancashire, United Kingdom}
\date{November 2021}

\begin{document}
\begin{abstract}
In this paper, we introduce mREAL-GAN, a generative adversarial network (GAN) for the parallel generation of multiple residential electrical appliance load (mREAL) profiles. mREAL-GAN is intended for use in community-scale low-voltage network analysis, and represents a departure from previous methods for this purpose, which break the generation of appliance load profiles into several steps and largely model each appliance independently. Instead, mREAL-GAN models appliance load profiles in an end-to-end manner, and generates multiple appliance load profiles in parallel in a way that captures inter-dependencies. We show that mREAL-GAN generates load profiles for individual appliance-types with greater fidelity than a popular example of previous methods, and demonstrate its ability to capture inter-dependencies between appliances.
\end{abstract}
\maketitle

\section{Introduction}

The rapid introduction of microgeneration technologies, transition to electric cars, and electrification of heating, present challenges for the operation of community-scale, low-voltage networks \cite{annex54}. Simulation-based analysis provides a means of assessing the impact of such developments, which can inform and coordinate strategies to ensure the integrity of the residential energy supply throughout. Simulating low-voltage networks for this purpose however requires an accurate representation of community-scale energy consumption, which captures a diverse range of realistic possibilities, and allows for the flexibility to consider the effect of potential future scenarios.

\subsection{Previous Work}\label{prevwork}
At the community-scale, in contrast to larger scales, energy consumption is highly influenced by outlier residences which deviate significantly from the group average \cite{flett}, as well as the fluctuations in demand caused by the switching on and off of individual devices \cite{richardson}. For these reasons, energy consumption models introduced for this purpose commonly take a bottom-up approach \cite{capasso}, where the load profile for each device in each residence is modelled first, prior to aggregation. In addition to reducing the difficulty of capturing the influence of outliers and the effect of devices switching on and off, in comparison to modelling community-scale energy consumption directly, this also has the advantage of allowing a user to explicitly include/exclude devices, providing greater flexibility. In this work, we focus on how load profiles for electrical appliances, the main subset of these devices, can be modelled. Existing bottom-up models of energy consumption can be classified, in terms of how they model load profiles for appliances, into four categories:
\begin{itemize}
\item \textit{Occupant-based models} \cite{walker1985,richardson,widen,muratori,wilke,alzate,fischer,diao,taniguchi}, which simulate occupant behaviour and infer features of appliance operating patterns (frequency, timing, duration, etc.) from this, before inferring load profiles from these patterns;
\item \textit{Empirical data-based models} \cite{paatero,armstrong,bajada,gruber,ortiz}, which simulate operating patterns based on empirical appliance load data and then also infer load profiles from these patterns;
\item \textit{Hybrid (occupant-based and empirical data-based) models} \cite{capasso,flett}, which simulate the frequency and duration of operations based on empirical data, and then infer the timing of operations from both these features and separately simulated occupant behaviour. Load profiles are then again inferred from the simulated operating patterns;
\item \textit{Other models} \cite{stokes,yao}, which take some alternative approach that cannot be classified into one of the other categories. These models were both published in 2005 and employed creative approaches to compensate for the lack of suitable data available at the time. To focus the discussion on more up-to-date techniques, these models will not be considered beyond this point.
\end{itemize}

While there is variation in these approaches, all break down the modelling of load profiles for appliances into several steps. Most notably, the majority of models explicitly simulate the operating patterns of appliances on some basis, before inferring load profiles directly from these patterns. This however assumes that operating patterns are the best information for inferring load profiles from, which, while seemingly intuitive, incorporates inductive bias that may constrain model performance. Additionally, empirical data-based models treat each appliance independently, when there are likely relationships between the operating patterns, and corresponding load profiles, for different appliances. For example, the operating pattern/load profile for a tumble dryer will likely have a strong relationship with the operating pattern/load profile for a washing machine in the same residence, since tumble dryers are generally used to dry clothes which have recently been washed in the washing machine. In one empirical data-based model \cite{armstrong}, the authors chose to address this by modelling the operating patterns of tumble dryers in a different manner to other appliances, which involves always inferring that a tumble dryer operation begins at some randomly sampled point within a certain window of time after a washing machine operation ends. However, this approach to capturing inter-dependencies between appliances is limiting for washing machines and tumble dryers, and is likely unsuitable for other groups of appliances that can be assumed to have closely related operating patterns. For washing machines and tumble dryers, the assumption constrains diversity. For example, the occupants of a residence who sometimes dry recently washed clothes in the tumble dryer may instead hang the clothes out to dry on a sunny day, which the approach would not allow for. For other groups of appliances that can be assumed to have related operating patterns/load profiles, the sequence of appliances operated can be assumed to vary, making it challenging to establish a similar principle. For example, consider the group of appliances commonly used in preparation of a meal, where different meals may involve operating appliances in a different order. With occupant-based and hybrid models, it can be said that inferring operating patterns from the same simulated occupant behaviour establishes some relationship between the generated load profiles for different appliances. However, this is likely still a weak representation of these relationships, since there is no direct consideration of the operating pattern of one appliance when inferring the operating pattern of another. Note, to allow for greater diversity, this inference of operating patterns from occupant behaviour is usually performed stochastically. Further to the points given here, other works \cite{mckennacritic, ramirez} have raised issues with the approaches used in occupant-based and hybrid models for simulating occupant behaviour and for inferring operating patterns, which are largely due to limitations established by the available data for human behaviour.

\subsection{Motivations and Contributions}
Our work explores the use of generative adversarial networks (GANs) \cite{goodfellow} for generating appliance load profiles on an end-to-end basis, i.e. without any intermediary steps. This was primarily motivated by the unprecedented performance GANs have shown in generating many forms of complex data, for example photographic images of human faces \cite{stylegan3}. Additionally, the end-to-end basis was considered promising as GANs use a deep neural network for the generation of a sample from a latent code, and deep neural networks trained to perform tasks in an end-to-end manner have been shown through a large number of developments since 2012 \cite{alexnet} to outperform modular approaches designed by humans in many cases \cite{dlreview}. We also identified a principle of image generating GANs that could allow for capturing the inter-dependencies between the load profiles of multiple appliances. This is that these models generate coloured images, which represent the colour of the image in each location (pixel) with three components corresponding to the intensity of red, green and blue (RGB). This establishes three channels of the image, which can be considered as separate but highly related parts of the same signal. It was realised that load profiles for multiple appliances in the same residence could similarly be represented as different channels of the same signal, and a GAN could be trained to generate these multi-channel signals, which would require capturing the inter-dependencies. We note that GANs have already been used in an appliance load model known as TraceGAN \cite{tracegan}, which is intended for use in the development of non-intrusive load monitoring (NILM) \cite{nilm} algorithms. However, TraceGAN differs from the framework introduced in this paper in that it generates a load profile for a single appliance at a time and therefore does not capture inter-dependencies, uses a conditional input for the user to specify which appliance-type out of five types to generate a load profile for, and uses a progressively growing GAN (ProGAN) architecture \cite{progan}. Additionally, TraceGAN is unsuitable for use in low-voltage network analysis where it is usually necessary to perform simulations of at least 24 hours in duration, e.g. \cite{montecarlostudy}, since the model is constrained to generating load profiles of 5 hours in duration.

\begin{figure}[t]
\makebox[\textwidth][c]{
	\includegraphics[width=12cm]{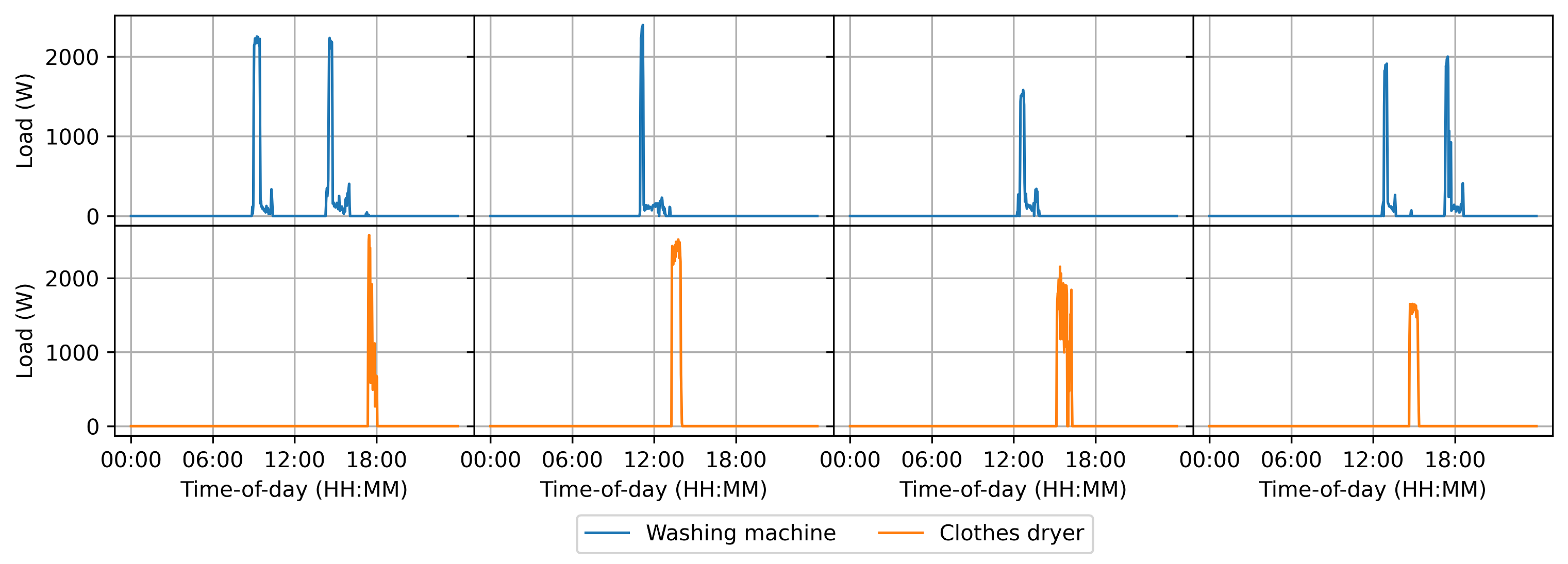}}
    \caption{Example samples of simultaneous 24-hour washing machine and tumble dryer load profiles, generated by mREAL-GAN. Each example was repeatedly randomly sampled until profiles exhibited an operation}
    \label{fig:examples}
\end{figure}

As the culmination of this work, this paper presents the multiple residential electrical appliance load GAN (mREAL-GAN), a framework for parallel generation of multiple appliance load profiles of 24-hours in duration. We show that for two appliances in the same residence (washing machines and tumble dryers), mREAL-GAN captures the inter-dependencies more accurately than a popular example of occupant-based models \cite{richardson}, which, as highlighted in Section \ref{prevwork}, relates appliances through a consistent representation of occupant behaviour for all appliances in the same residence. Additionally, it is shown that the individual load profiles generated by mREAL-GAN are of greater fidelity than those generated by the existing model. Some example pairs of washing machine and tumble dryer load profiles generated in parallel by mREAL-GAN are shown in Fig. \ref{fig:examples}. As can be seen from this visual perspective, the model generates samples which capture the general characteristics of the real data. Our findings confirm that an end-to-end GAN-based approach for modelling appliance load profiles both provides improvements over existing approaches, and allows for the additional functionality of capturing inter-dependencies between appliances. We also highlight further useful functionalities we believe could be incorporated in extension of this work and potential means of improving upon mREAL-GAN, based on the successes of GANs seen in other applications, as suggestions for future research.

\section{mREAL-GAN}
GANs are a form of latent variable generative model, where a generator $G$ is a function which maps a low-dimensional latent space described by a simple distribution $p(\boldsymbol{z})$, usually a standard Gaussian $\mathcal{N}(0,\boldsymbol{I})$, to an approximate for a manifold of high-dimensional real-world data embedded in $\mathbb{R}^d$. Samples of seemingly real-world data $\boldsymbol{x}_f\in \mathbb{R}^d$ may then be generated by first sampling in the latent space, before passing the sampled latent variable $\boldsymbol{z}\sim p(\boldsymbol{z})$ through the generator, i.e. $G(\boldsymbol{z})=\boldsymbol{x}_f$. Each possible value of the latent variable $\boldsymbol{z}$ therefore represents a latent code for each high-dimensional sample the generator can synthesise. The goal with latent variable generative models is then to find a generator function that best approximates the true manifold of real-world data.

A deep neural network is most typically used for the generator function in GANs, and is trained in a self-supervised manner through the use of a second model, known as the discriminator $D$. The discriminator and generator are trained simultaneously, with the discriminator being trained to map high-dimensional samples to a representation which maximises the statistical dissimilarity between the distributions of real and generated samples, and the generator being trained to generate samples that the discriminator would consider more realistic. By training both models simultaneously, the generator tries to incorporate the features of real-world data that the discriminator is using to distinguish real from generated samples at that time, with both models gradually improving their strategy until either converging on the optimal solution or the training is stopped. In contrast to the generator simply learning to generate the available examples of real data, this adversarial training procedure can allow generators to learn to synthesise samples of high-fidelity which are distinct from the training data.

\subsection{Loss Functions}\label{objective}
mREAL-GAN follows the described principle of GANs, and uses the Wasserstein-1 distance-based GAN objective \cite{wgan} for optimising $G$ and $D$. This involves training the discriminator to maximise the distance function $\tilde{W}$, while training the generator to minimise $\tilde{W}$, where:

\begin{equation}
\tilde{W}= \E_{\boldsymbol{x}_r\sim p\left(\boldsymbol{x}_r\right)}\left[D\left(\boldsymbol{x}_r\right)\right] - \E_{\boldsymbol{z}\sim p\left(\boldsymbol{z}\right)}\left[D\left(G\left(\boldsymbol{z}\right)\right)\right]
\end{equation}
\noindent where $\boldsymbol{x}_r\in \mathbb{R}^d$ is a real sample.

This distance function is based on the dual form of the Wasserstein-1 distance, which requires $D\in \mathcal{D}$, where $\mathcal{D}$ is the set of $K$-Lipschitz functions. We use the gradient penalty technique for enforcing the $K$-Lipschitz constraint \cite{wgangp}, where we set $K=1$, employing the distance function-scaled one-sided version of this penalty \cite{eeg}:

\begin{equation}\label{eq:gp2}
    P_g=\lambda\left(\max (0,\tilde{W})\right)
    \E_{\hat{\boldsymbol{x}} \sim p\left(\hat{\boldsymbol{x}}\right)}
    \left[\text{max}\left(0,\|\nabla_{\hat{\boldsymbol{x}}} D\left(\hat{\boldsymbol{x}}\right)\|_{2}-1\right)^{2}\right]
\end{equation}

\noindent where $\lambda$ is the weight of the gradient penalty, which we set as $\lambda=10$, and $\hat{\boldsymbol{x}}$ is a randomly sampled point along the straight line between a real sample and a generated sample:

\begin{equation}
\label{eq:sampleinterp}
\hat{\boldsymbol{x}}=\epsilon \boldsymbol{x}_r+\left(1-\epsilon\right) G\left(\boldsymbol{z}\right)
\end{equation}

\noindent where \(\epsilon\sim \mathcal{U}(0,1)\).

We also use the variant of the drift penalty \cite{progan} introduced in \cite{eeg} to centre the discriminator on zero:

\begin{equation}
P_c=\beta \left(\E_{\boldsymbol{x}_r \sim p\left(\boldsymbol{x}_r\right)} \left[D\left(\boldsymbol{x}_r\right)\right] + \E_{\boldsymbol{z} \sim p\left(\boldsymbol{z}\right)} \left[D\left(G\left(\boldsymbol{z}\right)\right)\right]\right)^2
\end{equation}

\noindent where $\beta$ is the weight of the drift penalty, which we set as $\beta=\text{1e-3}$.

This establishes the following loss functions for the generator and discriminator:

\begin{equation}
\mathcal{L}_G=\tilde{W}
\end{equation}

\begin{equation}
\mathcal{L}_D=-\tilde{W}+P_g+P_c
\end{equation}

Note, training the generator to minimise $\tilde{W}$ means minimising the term $-\E_{\boldsymbol{z}\sim p(\boldsymbol{z})}[D(G(\boldsymbol{z}))]$ with respect to $G$.

\subsection{Architecture}
\begin{figure}[tp]
    \centering
    \includegraphics[width=6.7cm]{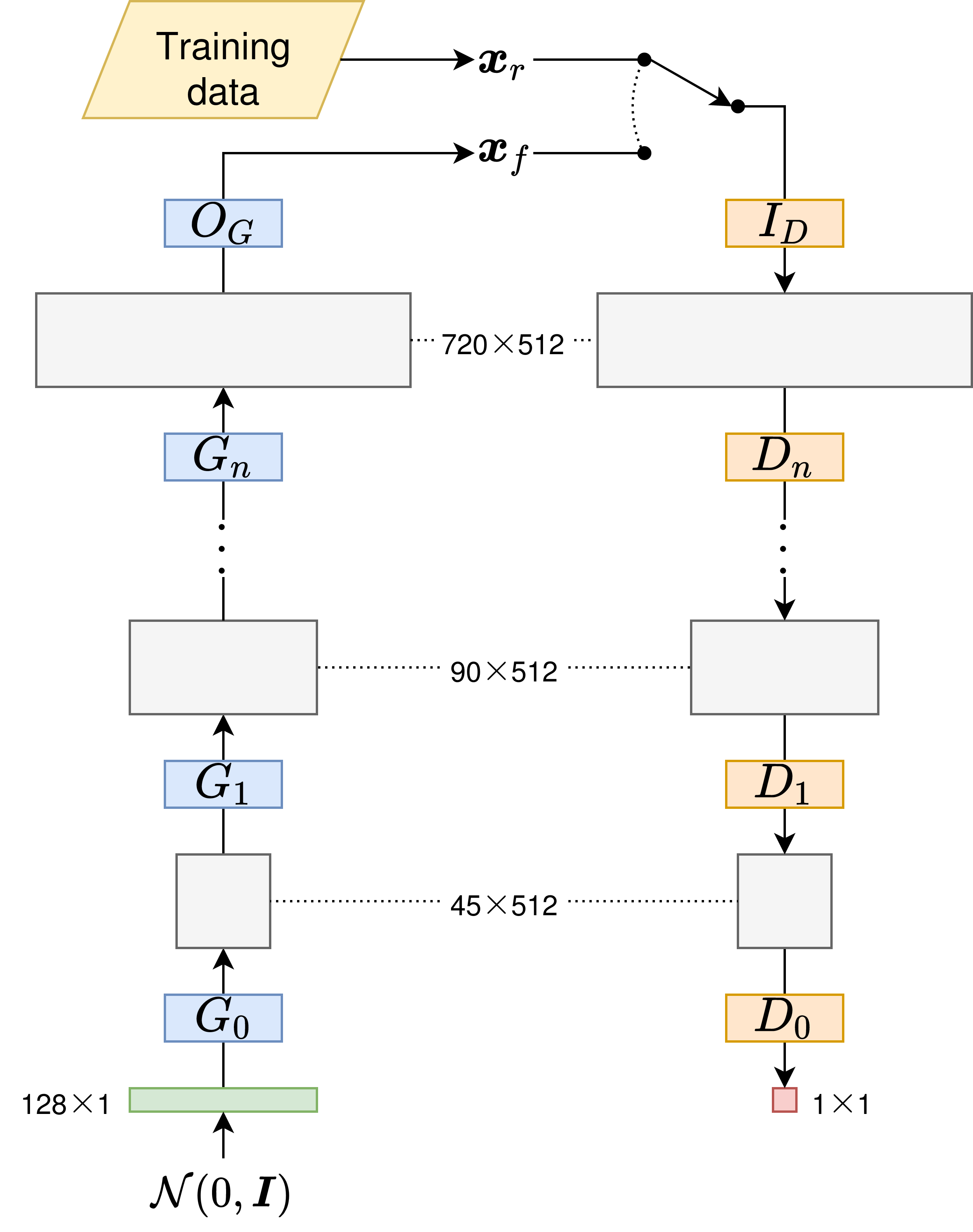}
    \caption{Schematic of our adaptation of the DCGAN architecture \cite{dcgan} used in mREAL-GAN}
    \label{fig:DCGAN SCHEMA}
\end{figure}

We used an adaptation of the robust DCGAN architecture \cite{dcgan}, which we show in Fig. \ref{fig:DCGAN SCHEMA}, in which the generator and the discriminator are both defined as compositions of functions, i.e.  \(G=O_G\circ G_n \circ \ldots G_1 \circ G_0\) and \(D=D_0\circ D_1 \circ \ldots D_n \circ I_D\), where we refer to $O_G$, $G_i$ for $i=0,1,\ldots,n$,  $D_i$ for $i=0,1,\ldots,n$, and $I_D$ as blocks. While more sophisticated architectures may be more suitable for our problem, this architecture provides a relatively reliable starting point when applying GANs to new forms of data, and an easily extendable structure for future work to build off.

The blocks in the generator are defined:

\begin{equation}
G_0=A_G\circ B_{G_0,3}\circ W_{G_0,2}\circ A_G\circ B_{G_0,2}\circ W_{G_0,1}\circ A_G\circ B_{G_0,1}\circ F_G
\end{equation}

\begin{equation}
G_i=A_G\circ B_{G_i,2}\circ W_{G_i,2}\circ A_G\circ B_{G_i,1}\circ W_{G_i,1}\circ R_{\uparrow},\quad i=1,2,\ldots,n
\end{equation}

\begin{equation}
O_G=A_{O_G}\circ W_{O_G}
\end{equation}

\noindent where \(A_G\) is the activation function used in the hidden layers of the generator, \(B_{G_i,j}\) is the \(j^{th}\) batch normalisation layer \cite{batchnorm} in \(G_i\), \(W_{G_i,j}\) is the \(j^{th}\) convolutional layer in \(G_i\), \(F_G\) is a fully connected layer, \(R_{\uparrow}\) upsamples its input to \(2\times\) the temporal resolution, \(A_{O_G}\) is the activation function used for the output layer, and \(W_{O_G}\) is the single convolutional layer in \(O_G\). Following the same convention, and defining \(R_{\downarrow}\) as downsampling the input to \(1/2\times\) the temporal resolution, the blocks of the discriminator are then as follows:

\begin{equation}
I_D=A_{D}\circ W_{I_D}
\end{equation}

\begin{equation}
D_i=R_{\downarrow}\circ A_D\circ W_{D_i,2}\circ A_D\circ W_{D_i,1},\quad i=1,2,\ldots,n
\end{equation}

\begin{equation}
D_0=F_D\circ A_D\circ W_{D_0,2}\circ A_D\circ W_{D_0,1}
\end{equation}

\subsubsection{Architectural Hyperparameters}\label{hyperparams}
We set $n=4$ and initialise the parameters of both models $\theta_G$ and  $\theta_D$ using Glorot initialisation \cite{glorot}. Then, starting at the input to the generator and ending at the output of the discriminator, the rest of the yet unspecified architectural hyperparameters are as follows. We define the fully connected layer in the generator as $F_G:\mathbb{R}^{128}\rightarrow \mathbb{R}^{45\times 512}$ (outputting 45 time-steps of 512 channels); use a Leaky ReLU with gradient for negative inputs $\alpha=0.2$ for the activation $A_G$; use 512 filters with 15 time-step kernels for the convolutional layer $W_{G_i,j}$, for $i=0,1,\ldots,n$ and $j=1,2$; use nearest neighbour upsampling for $R_{\uparrow}$; use 2 filters with 1 time-step kernels for the convolutional layer $W_{O_G}$; use a ReLU for $A_{O_G}$; use 512 filters with 1 time-step kernels for $W_{I_D}$;  use a Leaky ReLU with gradient $\alpha=0.2$ for negative inputs for the activation $A_D$; use 512 filters with 15 time-step kernels for the convolutional layer $W_{D_i,j}$, for $i=0,1,\ldots,n$ and $j=1,2$; use average pooling for $R_{\downarrow}$; and define the fully connected layer in the discriminator as $F_D:\mathbb{R}^{45\times 512}\rightarrow\mathbb{R}$.

The use of a ReLU for $A_{O_G}$ marks a departure from most work on GANs, which commonly use the tanh activation function instead. This decision was seen to provide a better representation of the switching on and off of appliances, where a sudden change in load from a constant 0W to a high Wattage or vice-versa is common. The tanh function instead does not allow the model to represent a constant 0W, and causes the load to ramp up and down at the beginning of operations.

\subsection{Training Procedure}\label{trainproc}
To train mREAL-GAN, we iterate Algorithm \ref{alg:train MDC} (see \ref{trainalgos}) for 100k steps. In this, as is common when using the Wasserstein-1 distance-based objective (see Section \ref{objective}), we update the discriminator several times before updating the generator a single time, setting the number of discriminator updates per train step $n_{dstep}=5$. We also use a minibatch size of $m=64$, apply hypersphere normalisation \cite{progan} to the latent vector, and use RMSProp \cite{rmsprop} for the updates with a learning rate of $\gamma=\text{1e-5}$. In addition, since our training data is of a relatively small number of samples, we introduce REAL-SDA, an adaptation of stochastic discriminator augmentation (SDA) \cite{ada} for load profile data based on random shifts in time and additive noise of a randomly sampled scale. The sampling of the time-shift and additive noise scale is shown in Algorithm \ref{alg:train MDC} and the augmentation in Algorithm \ref{alg:aug most} (see \ref{trainalgos}). We set the time-shift parameter $\rho=1024$ and the additive noise parameter $\eta=200$. For use in Algorithm \ref{alg:aug most}, we define the notational convention $\boldsymbol{x}_{a:b}=(x_a,x_{a+1},\ldots,x_b)$ for $a\le b$, and note that we use $\boldsymbol{x}\frown\boldsymbol{y}$ to denote concatenation of $\boldsymbol{x}$ and $\boldsymbol{y}$. Once training is completed, we take the exponential moving average (EMA) of the generators weights throughout training \cite{ema} to create the final generator, which we use for evaluation.

\section{Training Data}
We trained mREAL-GAN to generate simultaneous washing machine and tumble dryer load profiles, using data extracted from the cleaned version of the Household Electricity Survey (HES) \cite{hes}.  In collection of this data, sensors were fitted to appliances in 250 residences located around the UK to record electrical load, which took place between May 2010 and June 2011. In 224 of these residences, appliances were monitored at a high-temporal resolution of 2-minutes, for a period of between 20 and 45 days. To expand the dataset as much as possible with the available equipment, different residences were monitored over different periods.

For each residence and for every full day (midnight to midnight) the residence was monitored, we extracted the washing machine and tumble dryer load profile. Wherever an appliance was not monitored, usually due to not being owned by the residence, the appliance was represented with a load profile of a constant 0W. This established 6,215 samples (days) of data, where each sample consists of $720\times2=1440$ load values ($720\times2$-minute time-steps in a day for each of 2 appliances), with each load value corresponding to the average load over a particular 2-minute time-step of a day. We therefore define the number of appliances $n_{app}=2$, the number of 2-minute time-steps in a day $T=720$, and the dimensionality of the data $d=n_{app}T=1440$. To normalise the data for training, all load values for each appliance-type were divided by $6\sigma$, where $\sigma$ is the standard deviation of load values for that appliance-type. Note that when testing the tanh activation for $A_{O_G}$ (see Section \ref{hyperparams}), we instead normalised the load values to a range of $[-1, 1]$. Additionally, note that samples generated by mREAL-GAN were mapped back to the original scale of the measured data prior to any evaluation.

\section{Experiment}
For assessment of mREAL-GAN, we generated 6,215 samples (days) of simultaneous washing machine and tumble dryer load profiles and quantified the similarity of this data with the real data extracted from the HES \cite{hes}. We also did the same for a popular example of an existing appliance load model \cite{richardson}, which may be referred to as the Centre for Renewable Energy Systems Technology (CREST) model, to assess whether mREAL-GAN provided benefits. In generation of data for the CREST model, which takes the day-type (weekday or weekend) and the number of occupants in the residence as inputs, we ensured that each sample generated by the model corresponded, in terms of these conditions, to a real sample, on a one-to-one basis. However, since the model takes a maximum of 5 occupants, an input of 5 occupants was used for any residence which has a greater number of occupants. Additionally, since the model initialises with all appliances switched off, we took each sample as the second day from a two day simulation, and then resampled this from 1-minute to 2-minute resolution using average pooling.

It should be noted that assessing how well a model represents the data it was trained on in this way is the typical approach to evaluating the performance of generative models \cite{is,fid,progan,noteonevalofgenmodels}. However, this is usually done to compare models which have been trained on the same data and unlike mREAL-GAN, the CREST model \cite{richardson} was not trained on the HES data \cite{hes}. Addressing this here would mean either comparing mREAL-GAN against a model which has also been trained on the HES data, or using some separate dataset for performing the evaluation, however we could not identify any suitable examples of either option. For example, we considered the REFIT \cite{refit} dataset, but found that this suffers from a significant number of substantial gaps in the monitoring of each appliance, which makes extracting a consistent set of simultaneous 24-hour long load profiles for multiple appliances in the same residence challenging. Many other datasets were also deemed unsuitable on the basis that they cover different regions of the world, where both the HES data and the data used in development of the CREST model were collected in the UK. Additionally, we found no model also trained exclusively on the HES data. The comparison we present therefore assumes that the HES data is representative of the true distribution of UK load profiles. If this assumption is valid, it is reasonable to compare the models on this basis, since accurately representing this distribution is the general aim of both mREAL-GAN and the CREST model, regardless of which data were used in model development. Despite the uncertainty around this aspect of our experiment however, we note that future work can compare models also trained on the HES data against mREAL-GAN to reliably assess whether improvements have been made in representing this data.

We assessed how well both models represent the HES data by quantifying similarity in terms of four feature-types: the inter-dependencies between washing machines and tumble dryers; the distribution of load values for each appliance-type; the distribution of sub-sequences of load profiles for each appliance-type; and the distribution of load profiles for each appliance-type.

\subsection{Accuracy in Capturing the Inter-Dependencies Between Washing Machines and Tumble Dryers}
To assess how well the inter-dependencies between washing machines and tumble dryers, as described by the HES data, are captured by a model, we use the following test. This involves first computing the matrix $\boldsymbol{C}_L$ for a dataset $L$ of measured or generated samples (days):

\begin{equation}
\boldsymbol{C}_{L}=\begin{pmatrix}
c_{L,1,1} & c_{L,1,2} & \dots & c_{L,1,d} \\
c_{L,2,1} & c_{L,2,2} & \dots & c_{L,2,d} \\
\vdots & \vdots & \ddots & \vdots \\
c_{L,d,1} & c_{L,d,2} & \dots & c_{L,d,d}
\end{pmatrix}
\label{correlation}
\end{equation}

\noindent where $d=1440$, and $c_{L,i,j}$ is the correlation coefficient, a common measure of the dependence between two features, between the load value occurring at the $i^{th}$ time-step of the day for washing machines, and the load value occurring at the $j^{th}$ time-step of the day for tumble dryers, across all samples (days) in $L$.

Prior to computing $\boldsymbol{C}_L$ for a dataset $L$, due to some time-steps having a load of 0W across all samples in the measured load profiles for both washing machines and tumble dryers, we add Gaussian noise, sampling the value added to each load value independently from $\mathcal{N}(0,\text{1e-5})$. We then compute the correlation matrix distance \cite{correlationdist} between the $\boldsymbol{C}_L$ for real and generated datasets to assess the similarity.

We used this test with and without reducing the dimensionality of the individual appliance load profiles prior to computing the correlation matrix using average pooling, which had little effect on the measure. For simplicity, we present the results from applying the test without dimensionality reduction.

\subsection{Accuracy in Capturing the Distribution of Load Values for Each Appliance-Type}
The lowest-level feature of a load profile is a load value, i.e. the load for a single time-step. To measure the accuracy with which the distribution of load values for each appliance-type, as seen in the set of real data, is captured by a set of generated data, we first extract all load values from all load profiles in both sets of data for that appliance-type. This establishes $6215\times 720=4474800$ data points for each set. Then, we measure the similarity between the 1D distributions of load values for real and generated data using the Wasserstein-1 distance, which is generally more robust than other similarity measures \cite{bazan}, using the 1D point clouds to represent the distributions. Note that in the 1D case, the Wasserstein-1 distance has a closed-form solution \cite{swd}.

\subsection{Accuracy in Capturing the Distribution of Sub-Sequences of Load Profiles for Each Appliance-Type}
We may consider sub-sequences of load profiles as higher-level features than individual load values, but which are lower-level than the full load profiles. To assess how accurately the distribution of sub-sequences of load profiles for each appliance-type, as seen in the set of real data, is captured by a set of generated data, we first extract 32 sub-sequences of 45 time-steps in duration from each load profile in both sets of data for that appliance-type. The extraction is done on a uniformly random basis, and this establishes $6215\times 32=198880$ data points for each set. This extraction is inspired by the approach used to evaluate generated images in \cite{progan}, where patches are randomly extracted instead of sub-sequences. In contrast to \cite{progan} however, we do not normalise the resulting sets of data points, since the magnitude of features (load values) is of significance here. We then measure the similarity between the 45D distributions of sub-sequences for real and generated data using the sliced Wasserstein distance (SWD) \cite{swd}, using the 45D point clouds to represent the distributions. SWD is a popular alternative to the Wasserstein-1 distance, which uses random 1D projections of the point clouds and the closed-form solution of the Wasserstein-1 distance in 1D to provide an efficient measure of similarity between multi-dimensional point clouds but which incorporates the key benefits of the Wasserstein-1 distance. For this test, we used 128 projections.

\subsection{Accuracy in Capturing the Distribution of Load Profiles for Each Appliance-Type}
The final test we consider assesses the accuracy with which the distribution of load profiles for each appliance-type, as seen in the set of real data, is captured by a set of generated data. Based on the load profiles having a dimensionality of 720, the number of samples (6215) is relatively low for such an assessment. As such, we first reduce the dimensionality of the load profiles to 144 using average pooling with a pool size of 5 time-steps. We then measure the similarity between the 720D distributions of real and generated load profiles using the sliced Wasserstein distance (SWD) \cite{swd}, using the 144D point clouds to represent the distributions, and using 512 projections.

\section{Results}
The results from our experiment are given in Table \ref{tab:results}, where a lower value indicates better similarity for all measures. From this, it can be seen that mREAL-GAN outperforms the CREST model in all tests of performance.

\begin{table}[tp]
\makebox[\textwidth][c]{
\begin{tabular}{?c~c~ccc~ccc?}
\hlinewd{0.9pt}
          &                                                               & \multicolumn{3}{c~}{Washing machines}                                                                                                                                      & \multicolumn{3}{c?}{Tumble dryers}                                                                                                                                          \\
          & \begin{tabular}[c]{@{}c@{}}Inter-\\ dependencies\end{tabular} & \begin{tabular}[c]{@{}c@{}}Load\\ values\end{tabular} & \begin{tabular}[c]{@{}c@{}}Sub-\\ sequences\end{tabular} & \begin{tabular}[c]{@{}c@{}}Load\\ profiles\end{tabular} & \begin{tabular}[c]{@{}c@{}}Load\\ values\end{tabular} & \begin{tabular}[c]{@{}c@{}}Sub-\\ sequences\end{tabular} & \begin{tabular}[c]{@{}c@{}}Load\\ profiles\end{tabular} \\ \hlinewd{0.7pt}
mREAL-GAN & 0.39                                                          & \multicolumn{1}{c|}{2.11}                             & \multicolumn{1}{c|}{3.61}                                & 45.07                                                   & \multicolumn{1}{c|}{3.85}                             & \multicolumn{1}{c|}{4.52}                                & 37.35                                                   \\ \hline
CREST \cite{richardson}     & 0.64                                                          & \multicolumn{1}{c|}{8.13}                             & \multicolumn{1}{c|}{9.93}                                & 78.43                                                   & \multicolumn{1}{c|}{21.23}                            & \multicolumn{1}{c|}{27.00}                               & 119.26                                                  \\ \hlinewd{0.9pt}
\end{tabular}}
\caption{Results from our testing of the similarity between generated and measured data, in terms of the inter-dependencies between appliance-types, and for each appliance-type: the distribution of load values, the distribution of sub-sequences of load profiles, and the distribution of load profiles.}\label{tab:results}
\end{table}

Notably, based on the results for the generation of load profiles for both appliance-types, the CREST model generates washing machine load profiles which are more comparable to mREAL-GAN in terms of similarity to the HES data than it does tumble dryer load profiles. This is likely due to the representation of load throughout operation in the CREST model being more realistic in the case of washing machines, where the model uses a typical load throughout operation curve to infer a load profile from an operating pattern (see Section \ref{prevwork}). Instead, the load throughout operation for tumble dryers is represented as a constant value. The approach used for washing machines can however be considered an outlier, as the model represents the load throughout operation for all other appliances, similarly to how most models treat all appliances including washing machines, with a constant value. It can therefore be assumed that mREAL-GAN would likely outperform these other models in equivalent tests. Even with the better representation for washing machines in the CREST model however, the approach assumes that this load throughout operation curve does not vary between operations of the same appliance, or between appliances. This greatly reduces diversity, since these curves can vary for the same washing machine due to different modes of operation, as well as variation in environmental conditions, and the curve can vary between different washing machines due to different make/model and age. The ability of GAN-based models to generate load profiles which allow for these variations is therefore of significant benefit.

\section{Future work}
We envisage a wide variety of ways this work could be improved on and extended from in future work. To improve upon mREAL-GAN in the task we have demonstrated its performance in, we recommend testing alternative GAN architectures, or other forms of deep generative model such as variational autoencoders (VAEs) \cite{vae}, flow-based models \cite{fbm}, or denoising diffusion probabilistic models \cite{denoising}, the latter of which have been recently shown to outperform GANs on some of the most challenging generative modelling benchmarks \cite{denoisingbeatsgans}. Additionally, the techniques used in mREAL-GAN could likely be improved on, for example an adaptation of adaptive discriminator augmentation (ADA) \cite{ada} would likely provide benefits over REAL-SDA used in this work (see Section \ref{trainproc}).

Beyond this task, there is much potential for applying mREAL-GAN, or comparable techniques, to other load profile data, most notably for other regions of the world and for other groups of appliances. There is then a great deal of possible conditional functionalities that could be incorporated to allow a user to generate samples with particular characteristics, which could be beneficial in this area. For example, a user may want the functionality to generate samples for a residence with particular characteristics. This form of conditional functionality has already been introduced for GANs \cite{cgan}. More specific conditional functionalities may also be useful, for example the ability to generate a load profile where an operation occurs over a particular sub-sequence of time-steps, similar to how \cite{tumorthing} generated medical images with tumors in user-defined regions of the image. This would likely greatly aid in the simulation-based analysis of demand side management (DSM) strategies.

\section{Conclusion}
In this paper, we presented mREAL-GAN, a framework for parallel generation of multiple residential electrical appliance load profiles. We showed that the framework allows for capturing the inter-dependencies between appliance-types, a functionality which has not previously been addressed, as well as a more accurate representation of the distributions of load profiles for individual appliance-types. Our work establishes the foundations for further work on adapting deep generative modelling techniques to this problem, which should allow for new model strengths and functionalities to better aid the simulation-based analysis of community-scale low-voltage networks.

\bibliographystyle{unsrt}
\bibliography{Paper.bib}
\appendix
\newpage
\section{Training algorithms}\label{trainalgos}

\begin{algorithm}
\caption{Augmentation of an individual appliance load profile used in REAL-SDA}
\label{alg:aug most}
\begin{algorithmic}
\REQUIRE Real/generated sample $\boldsymbol{x}$, Time-shift $\Delta$, Additive noise scale $\sigma$
\STATE Sample noise vector $\boldsymbol{q}\in\mathbb{R}^{T}\sim \mathcal{N}(0,\sigma^2\boldsymbol{I})$
\IF{$\Delta=0$}
\STATE Add noise $\boldsymbol{x}\leftarrow\boldsymbol{x}+\boldsymbol{q}$
\ELSIF{$\Delta>0$}
\STATE Shift and add noise $\boldsymbol{x}\leftarrow \boldsymbol{x}_{\Delta+1:T}\frown\boldsymbol{x}_{1:\Delta}+\boldsymbol{q}$
\ELSE
\STATE Shift and add noise $\boldsymbol{x}\leftarrow \boldsymbol{x}_{T-\Delta+1:T}\frown\boldsymbol{x}_{1:T-\Delta}+\boldsymbol{q}$
\ENDIF
\ENSURE Augmented sample $\boldsymbol{x}$
\end{algorithmic}
\end{algorithm}

\begin{algorithm}[htp]
\caption{Train step for mREAL-DCGAN}
\label{alg:train MDC}
  \small
\begin{algorithmic}
\REQUIRE Model parameters $\theta_G$, $\theta_D$, Minibatch size $m$, Additive noise parameter $\eta$, Time-shift parameter $\rho$, Gradient penalty weight $\lambda$, Drift penalty weight $\beta$, Learning rate $\gamma$, Training data $L_{train}$, Number of appliances $n_{app}$, Number of discriminator steps $n_{dstep}$
\STATE Sample $\mu\sim\mathcal{N}(0,\rho)$
\STATE Get time-shift $\Delta\leftarrow \lfloor\mu\rfloor$
\FOR{$n_{dstep}$ iterations}
\FOR{$i=1,2,\ldots,m$}
\STATE Get real sample \(\boldsymbol{x}_{r,i}\) from $L_{train}$
\STATE Sample latent vector \(\boldsymbol{z}\sim \mathcal{N}(0,\boldsymbol{I})\)
\STATE Normalise latent vector $\boldsymbol{z}\leftarrow\boldsymbol{z}/||\boldsymbol{z}||_2$
\STATE Get fake sample $\boldsymbol{x}_{f,i}\leftarrow G(\boldsymbol{z})$
\STATE Sample additive noise scale $\sigma\sim e(\eta)$
\FOR{$j=1,2,\ldots,n_{app}$}
\STATE Extract $j^{th}$ individual appliance load profile $\boldsymbol{x}_{r,i,j}$ from $\boldsymbol{x}_{r,i}$
\STATE Augment load profile $\boldsymbol{x}_{r,i,j}\leftarrow \text{Aug}(\boldsymbol{x}_{r,i,j},\Delta,\sigma)$ (Algorithm \ref{alg:aug most})
\ENDFOR
\STATE Concatenate $\boldsymbol{x}_{r,i}\leftarrow \boldsymbol{x}_{r,i,1}\frown\boldsymbol{x}_{r,i,2}\frown\ldots\frown\boldsymbol{x}_{r,i,n_{app}}$
\STATE Sample additive noise scale $\sigma\sim e(\eta)$
\FOR{$j=1,2,\ldots,n_{app}$}
\STATE Extract $j^{th}$ individual appliance load profile $\boldsymbol{x}_{f,i,j}$ from $\boldsymbol{x}_{f,i}$
\STATE Augment load profile $\boldsymbol{x}_{f,i,j}\leftarrow \text{Aug}(\boldsymbol{x}_{f,i,j},\Delta,\sigma)$ (Algorithm \ref{alg:aug most})
\ENDFOR
\STATE Concatenate $\boldsymbol{x}_{f,i}\leftarrow \boldsymbol{x}_{f,i,1}\frown\boldsymbol{x}_{f,i,2}\frown\ldots\frown\boldsymbol{x}_{f,i,n_{app}}$ 
\STATE Sample $\epsilon\sim \mathcal{U}(0,1)$
\STATE Get interpolated sample $\hat{\boldsymbol{x}}_i\leftarrow\epsilon \boldsymbol{x}_{r,i}+(1-\epsilon)\boldsymbol{x}_{f,i}$
\ENDFOR
\STATE Get discriminator objective penalty $P_D\leftarrow -\frac{1}{m}\sum_{i=1}^{m}D(\boldsymbol{x}_{r,i})+\frac{1}{m}\sum_{i=1}^{m}D(\boldsymbol{x}_{f,i})$
\STATE Get gradient penalty $P_g\leftarrow\lambda(\max(0,-P_D))\frac{1}{m}\sum_{i=1}^{m}\max(0,||\nabla_{\hat{\boldsymbol{x}}_i}D(\hat{\boldsymbol{x}}_i)||_2-1)^2$
\STATE Get drift penalty $P_c\leftarrow \beta(\frac{1}{m}\sum_{i=1}^{m}D(\boldsymbol{x}_{r,i})+ \frac{1}{m}\sum_{i=1}^{m}D(\boldsymbol{x}_{f,i}))$
\STATE Update discriminator $\theta_D \leftarrow \theta_D - \gamma\text{RMSProp}(\theta_D,\nabla_{\theta_D}(P_D+P_g+P_c))$
\ENDFOR
\FOR{$i=1,2,\ldots,m$}
\STATE Sample latent vector \(\boldsymbol{z}\sim \mathcal{N}(0,1)\)
\STATE Normalise latent vector $\boldsymbol{z}\leftarrow\boldsymbol{z}/||\boldsymbol{z}||_2$
\STATE Get fake sample $\boldsymbol{x}_{f,i}\leftarrow G(\boldsymbol{z})$
\STATE Sample additive noise scale $\sigma\sim e(\eta)$
\FOR{$j=1,2,\ldots,n_{app}$}
\STATE Extract $j^{th}$ individual appliance load profile $\boldsymbol{x}_{f,i,j}$ from $\boldsymbol{x}_{f,i}$
\STATE Augment load profile $\boldsymbol{x}_{f,i,j}\leftarrow \text{Aug}(\boldsymbol{x}_{f,i,j},\Delta,\sigma)$ (Algorithm \ref{alg:aug most})
\ENDFOR
\STATE Concatenate $\boldsymbol{x}_{f,i}\leftarrow \boldsymbol{x}_{f,i,1}\frown\boldsymbol{x}_{f,i,2}\frown \ldots\frown\boldsymbol{x}_{f,i,n_{app}}$
\ENDFOR
\STATE Get generator objective penalty $P_G \leftarrow -\frac{1}{m}\sum_{i=1}^{m}D(\boldsymbol{x}_{f,i})$
\STATE Update generator $\theta_G \leftarrow \theta_G - \gamma\text{RMSProp}(\theta_G,\nabla_{\theta_G}(P_G))$
\ENSURE Model parameters $\theta_G$, $\theta_D$
\end{algorithmic}
\end{algorithm}

\end{document}